 \documentstyle[12pt]{article}  

\let\a=\alpha    \let\e=\epsilon
    
   \let\x=\xi \let\p=\pi

\let\la=\label  
   
\def\nn{\nonumber} \def\bd{\begin{document}} \def\ed{\end{document}}
\def\ds{\documentstyle} \let\fr=\frac \let\bl=\bigl \let\br=\bigr
\let\Br=\Bigr \let\Bl=\Bigl 
\let\bm=\bibitem
\let\na=\nabla
\let\pa=\partial \let\ov=\overline 
\newcommand{\be}{\begin{equation}} 
\newcommand{\ee}{\end{equation}} 
\def\ba{\begin{array}}
\def\ea{\end{array}}
\newcommand{\ho}[1]{$\, ^{#1}$}
\newcommand{\hoch}[1]{$\, ^{#1}$}
\newcommand{\bea}{\begin{eqnarray}} 
\newcommand{\eea}{\end{eqnarray}} 
\newcommand{\ra}{\rightarrow}
\newcommand{\lra}{\longrightarrow}
\newcommand{\Lra}{\Leftrightarrow}
\newcommand{\ap}{\alpha^\prime}
\newcommand{\bp}{\tilde \beta^\prime}
\newcommand{\tr}{{\rm tr} }
\newcommand{\Tr}{{\rm Tr} } 
\newcommand{\NP}{Nucl. Phys. }
\newcommand{\tamphys}{\it Isaac Newton Institute for Mathematical Sciences\\ University of Cambridge
 \\ 20 Clarkson Road, Cambridge CB3 OEH, U.K.}

\newcommand{\auth}{M. J. Duff \footnote{On leave of absence from the Center for Theoretical Physics, Texas A\&M University,
College Station, Texas 77843. Research supported in part by NSF Grant PHY-9411543.}}

\begin{document}
\hfill{NI-94-033}

\hfill{CTP-TAMU-49/94}

\hfill{hep-th/9501030}

\vspace{24pt}

\begin{center}
{ \large {\bf STRONG/WEAK COUPLING DUALITY FROM THE DUAL STRING }}

\vspace{36pt}

\auth

\vspace{10pt}

{\tamphys}

\vspace{44pt}

\underline{ABSTRACT}

\end{center}

On compactification to six spacetime dimensions, the fundamental heterotic string admits as a
soliton a dual string whose effective worldsheet action couples to the background fields of the dual
formulation of six-dimensional supergravity. On further toroidal compactification to four spacetime
dimensions, the dual string acquires an $O(2,2;Z)$ target-space duality.  This contains as a
subgroup the axion-dilaton $SL(2,Z)$ which corresponds to a strong/weak coupling duality for the
fundamental string.  The dual string also provides a new non-perturbative mechanism for
enhancement of the gauge symmetry.

{\vfill\leftline{}\vfill
\leftline{December 1994}
\pagebreak

\section{Introduction}
\la{Introduction}

It is becoming apparent that we must face up to {\it non-perturbative} effects in string theory
if we are to answer many of its important questions.  In this context, a fascinating
new idea is that of $S$-duality, according to which the four-dimensional heterotic string
compactified on a generic torus exhibits an $SL(2,Z)$ invariance: 
\be
S \rightarrow \frac{aS+b}{cS+d}
\la{sl2zs}
\ee  
where $a,b,c,d$ are integers satisfying $ad-bc=1$, where
\be
S=S_1+iS_2={\rm a}+ie^{-\eta}
\la{S}
\ee
and where ${\rm a}$ and $\eta$ are the four-dimensional axion and dilaton fields.  When accompanied
by electric/magnetic duality transformations on the gauge fields, the transformation (\ref{sl2zs}) is
known \cite{Cremmer} to be a symmetry of the field theory limit (which is given by $N=4$ supergravity
coupled to $22$ vector multiplets), and has been conjectured to be a symmetry of the full string
theory \cite{Font,Rey}.  See also \cite{Lu1}, where it was conjectured that discrete subgroups of all
the old non-compact global symmetries of compactified supergravity \cite{Julia,Marcus} (e.g
$SL(2,R)$, $O(6,22)$, $O(8,24)$, $E_7$, $E_8$, $E_9$, $E_{10}$) should be promoted to duality
symmetries of either heterotic or Type II superstrings\footnote{This was motivated partly by the
observation that the $D=11$ supermembrane \cite{Bergshoeff}, whose worldvolume couples to the
background fields of $D=11$ supergravity, treats the dilaton and the moduli fields on the same
footing.  The significance of this observation for string theory remains obscure.}.  We shall refer
to the transformation (\ref{sl2zs}) as $SL(2,Z)_S$.  Such a symmetry would be inherently
non-perturbative since it contains, in particular, the strong/weak coupling duality    %
\be  
{\rm g}^2/2\pi \rightarrow2\pi/{\rm g}^2  
\la{g} 
\ee
where ${\rm g}$ is the string loop expansion parameter given by 
\be
{\rm
g}^2/2\pi=<e^{\eta}>=8G/\alpha'
\ee
Here $G$ is Newton's constant and $2\pi\alpha'$ is the inverse
string tension. $S$-duality thus provides a stringy version of the Montonen-Olive
electric/magnetic conjecture \cite{Goddard,Montonen,Olive,Osborn} in globally supersymmetric
Yang-Mills theories, suitably generalized to include the $\theta$
angle \cite{Cardy1,Cardy2,Shapere,Sen1,Girardello,Vafa,Seiberg1,Seiberg2}.  More
recent evidence for $S$-duality in string theory may be found in
\cite{Kalara,Sen2,Sen3,Sen4,Sen5,Schwarz1,Schwarz2,Binetruy,Khuri1,Khuri2,Sen6,Rahmfeld,Harvey,Sen7,Hull,Duff,Khuri4}.

There is a formal similarity between this symmetry and
that of $T$-duality  \cite{Kikkawa,Sakai,Buscher,Nair,Cecotti,Giveon,Duff1} which acts on the
moduli fields. In fact, if we focus just on the moduli that arise in compactification on a
$2$-torus, the $T$-duality is just $O(2,2;Z)$:  
\be
M\rightarrow \Omega M  \Omega^T
\la{o22z}
\ee
where $\Omega$ is an O(2,2;Z) matrix satisfyling $\Omega^TL\Omega=L$, $M$ is a $4 \times 4$
matrix satisfying  $M=M^T$ and $MLM^T=L$ where  
\be
L=\left( \begin{array}{cc}
0&I_2\\
I_2&0
\end{array}
\right)
\ee
and where $I_2$ is the $2 \times 2$ identity matrix.  Explicitly,
\be
M=\left( \begin{array}{cc}
G^{-1}&G^{-1}B\\
-BG^{-1}&G-BG^{-1}B
\end{array}\right)
\la{M}
\ee
where $G$ and $B$ refer to the string metric $G_{mn}$ and 2-form $B_{mn}$ and, denoting the 
coordinates of the torus by $x^4,x^5$, where $m,n=4,5$.
 
In fact this contains as a subgroup another $SL(2,Z)$ transformation:
\be
T \rightarrow \frac{eT+f}{gT+h}
\la{sl2zt}
\ee  
where $e,f,g,h$ are integers satisfying $eg-fh=1$ and where
\be
T=T_1+iT_2={\rm b}+ie^{-\sigma}=B_{45} + i \sqrt{det G_{mn}} 
\la{T}
\ee
We shall refer to the transformation (\ref{sl2zt}) as $SL(2,Z)_T$.  Thus this $T$-duality may be
written as
\be
O(2,2;Z)_{T} \sim SL(2,Z)_O \times SL(2,Z)_T
\la{Tsplit}
\ee
where the $SL(2,Z)_O$ acts trivially.
In contrast to $SL(2,Z)_S$, $SL(2,Z)_T$ is known to be not merely a symmetry of the supergravity
theory but an exact string symmetry order by order in string perturbation
theory. It does, however, contain a minimum/maximum length duality mathematically similar to 
(\ref{g})  
\be
R \rightarrow \alpha'/R
\la{R}
\ee
where $R$ is the compactification scale given by
\be
\alpha'/R^2=<e^{\sigma}>.
\ee
We shall argue in this paper that these mathematical similarities between 
$SL(2,Z)_S$ and $SL(2,Z)_T$ are not coincidental. We shall present evidence in favor of the idea
that the physics of the fundamental heterotic string in six spacetime dimensions may equally
well be described by a dual heterotic string that emerges as a soliton solution of the fundamental
string \cite{Lu2,Lu3,Minasian}.  The inverse tension of the dual string, $2\pi \tilde
\alpha'$, is related to that of the fundamental string by the Dirac quantization
rule \cite{Lu3}
\be 
2\kappa{}^2=n(2\pi)^3\alpha'\tilde \alpha'\,\,\,\,\,\,n=integer
\la{Dirac1}
\ee
where $\kappa$ is the six-dimensional gravitational constant. Moreover, just as the fundamental
string worldsheet couples to the background fields of six-dimensional supergravity, so the dual
string worldsheet couples to the dual formulation of six-dimensional supergravity for which, in
particular, the six-dimensional dilaton $\Phi$ is replaced by $-\Phi$ and for which the 3-form
field strength $H=dB+...$ is replaced by its dual $\tilde H=d\tilde B+...$ where
$\tilde H=e^{-\Phi}*H$. This implies that the fundamental string metric $G_{MN}$ and the dual
string metric $\tilde G_{MN}$ $(M,N=0,...,5)$ are related to the canonical metric $G^c{}_{MN}$ by
$e^{-\Phi/2}G_{MN}=G^c{}_{MN}=e^{\Phi/2} \tilde G_{MN}$.  Since the dilaton enters the dual string
equations with the opposite sign to the fundamental string, it was argued in
\cite{Lu2,Lu3,Minasian} that in $D=6$ the strong coupling regime of the string should correspond to
the weak coupling regime of the dual string: 
\be
{\rm g}_6{}^2/(2\pi)^3 = <e^{\Phi/2}>=(2\pi)^3/{\tilde{\rm g}_6}^2
\la{coupling}
\ee
where ${\rm g_6}$ and $\tilde{\rm g}_6$ are the six-dimensional string and dual string
loop expansion parameters.

On compactification to four spacetime dimensions, the two theories appear very similar,
each acquiring an $O(2,2;Z)$ target space duality.  One's
first guess might be to assume that the strongly coupled four-dimensional fundamental
string corresponds to the weakly coupled dual string, but in fact something more subtle and
interesting happens: the roles of the $S$ and $T$ fields are interchanged \cite{Khuri1} so that the
strong/weak coupling $SL(2,Z)_S$ of the fundamental string emerges as subgroup of the target space
duality of the dual string:   
\be
O(2,2;Z)_{S} \sim SL(2,Z)_O \times SL(2,Z)_S
\la{Ssplit}
\ee
This {\it duality of dualities} is summarized in Table \ref{Table1}.           
\begin{table}
$
\begin{array}{lll}
&Fundamental \, string&Dual \, string\\
&&\\
T-duality&O(2,2;Z)_T &O(2,2;Z)_S\\
&\sim SL(2,Z)_O \times  SL(2,Z)_T&\sim SL(2,Z)_O\times  SL(2,Z)_S\\
Moduli&T={\rm b}+ie^{-\sigma}&S={\rm a}+ie^{-\eta}\\
&{\rm b}=B_{45}&{\rm a}=\tilde B_{45}\\
&e^{-\sigma}=\sqrt{detG_{mn}}&e^{-\eta}=\sqrt{det \tilde{G}_{mn}}\\
Worldsheet \, coupling&<e^{\sigma}>=\alpha'/R^2&<e^{\eta}>={\rm g}^2/2\pi\\
Large/small \, radius &R\rightarrow \alpha'/R&{\rm g}^2/2\pi\rightarrow 2\pi/{\rm g}^2\\
S-duality&SL(2,Z)_S&SL(2,Z)_T\\
Axion/dilaton&S={\rm a}+ie^{-\eta}&T={\rm b}+ie^{-\sigma}\\
&d{\rm a}=e^{-\eta}*H&d{\rm b}=e^{-\sigma}\tilde{*} \tilde{H}\\
&e^{-\eta}=e^{-\Phi}\sqrt{detG_{mn}}&e^{-\sigma}=e^{\Phi}\sqrt{det \tilde{G}_{mn}}\\
Spacetime \, coupling&<e^{\eta}>={\rm g}^2/2\pi&<e^{\sigma}>=\alpha'/R^2\\
Strong/weak \, coupling&{\rm g}^2/2\pi\rightarrow 2\pi/{\rm g}^2&R\rightarrow \alpha'/R
\end{array}
$
\label{Table1}
\caption{Duality of dualities}
\end{table}
As a consistency check, we note that since $(2\pi R)^2/2\kappa^2=1/16\pi G$ the Dirac
quantization rule (\ref{Dirac1}) becomes (choosing $n$=1)
\be 
8GR^2=\alpha'\tilde \alpha'
\la{Dirac2}
\ee
Invariance of this rule now requires that a strong/weak coupling transformation on the fundamental
string ($8G/\alpha'\rightarrow \alpha'/8G$) must be accompanied by a minimum/maximum length
transformation of the dual string ($\tilde \alpha'/R^2 \rightarrow R^2/\tilde \alpha'$), and vice
versa.

String theory requires two kinds of loop expansion: classical
$(\alpha')$ worldsheet loops with expansion parameter $<e^{\sigma}>$ and
quantum $(\hbar)$ spacetime loops with expansion parameter $<e^{\eta}>$. 
Since four-dimensional string/string duality interchanges the roles of $S={\rm a}+ie^{-\eta}$ and
$T={\rm b}+ie^{-\sigma}$, it also interchanges the roles of classical and quantum \cite{Duff}. So just
as we may write 
\be 
<e^{\eta}>=g^2=8G/\alpha'=\tilde \alpha'/R^2 
\ee
where $g$ is the fundamental string spacetime loop expansion parameter, so we may also write
\be 
<e^{\sigma}>=\tilde g^2=8G/\tilde \alpha'=\alpha'/R^2 
\ee 
where $\tilde g$ is the dual string spacetime loop expansion parameter.

This picture is similar in spirit to, though different in detail from, the conjecture of
Schwarz and Sen \cite{Schwarz1,Sen7} that $SL(2,Z)_S$ corresponds to a target space duality of a
toroidally compactified {\it fivebrane}\footnote{ See also \cite{Lu1}, where it was
pointed out that, on compactification from $D=10$ to $D=2$, the number of string moduli (given by
$G_{mn}$ and $B_{mn}$; m,n=1,...,8) is equal to the number of fivebrane moduli (given by $\tilde
G_{mn}$ and $\tilde B_{mnpqrs}$) and hence that both might exhibit the same $O(8,8;Z)$
target-space duality (or $O(8,24;Z)$ if we include the heterotic degrees of freedom) thus
combining $S$ and $T$.}; an idea based
on the earlier conjecture of a $D=10$ {\it string/fivebrane duality}\footnote{Ten-dimensional
string/fivebrane duality also interchanges the loop expansions \cite{Lu2,Lu3}.}
\cite{Duff2,Strominger,Lu1,Lu2,Lu3,Lu4,Lu5,Callan1,Callan2,Khuri3,Dixon,Khuri4}. We shall return to
this in a future publication but for the moment note that the $D=6$ {\it string/string duality}
picture presented here avoids the thorny problem of how to quantize a fundamental fivebrane.

\section{Clues from supergravity}
\la{supergravity}

Before discussing the solitonic string and determining its properties, we first wish to review
some properties of $D=6$ supergravity \cite{Salam} that already provide a clue to the
above-mentioned duality of dualities. There are two
formulations of six-dimensional supergravity, both with a $3$-form field strength\footnote{There
is also a {\it self-dual} supergravity in $D=6$ \cite{Salam} with a self-dual field strength and an
associated self-dual string \cite{Lu3}.  While interesting in its own right, this is not the
subject of the present paper.}. Let us denote the spacetime indices by $(M,N=0,...,5)$. Then the
bosonic part of the usual action takes the form      
\[ I_6=\frac{1}{2\kappa^2}\int d^6x \sqrt{-G}e^{-\Phi}[R_G+G^{MN}\partial_M\Phi\partial_N\Phi \]
\be
-\frac{1}{12}G^{MQ}G^{NR}G^{PS}H_{MNP}H_{QRS}+...] 
\la{a}
\ee
To within Chern--Simons corrections, $H$ is the curl of a 2-form $B$ 
\be
H=dB+...
\la{b}
\ee
The metric $G_{MN}$ is related to the canonical Einstein metric $G^c{}_{MN}$
by
\be
G_{MN}=e^{\Phi/2}G^c{}_{MN}
\la{metric}
\ee
where $\Phi$ the $D=6$ dilaton. Similarly, the dual supergravity action is given by 
\[
\tilde I_6=\frac{1}{2\kappa^2}\int d^6x \sqrt {-\tilde G}e^{\Phi}[
R_{\tilde G}+\tilde G^{MN}\partial_M\Phi\partial_N\Phi \]
\be
-\frac{1}{12}\tilde G^{MQ}\tilde G^{NR}\tilde G^{PS}\tilde H_{MNP}\tilde H_{QRS}+...] 
\la{c}
\ee
To within Chern-Simons corrections, $\tilde H$ is also the curl of a 2-form $\tilde B$ 
\be
\tilde H=d\tilde B+...
\la{d}
\ee
The dual metric $\tilde G_{MN}$ is related to the canonical Einstein metric by
\be
\tilde G_{MN}=e^{-\Phi/2}G^c{}_{MN}
\la{e}
\ee
The two supergravities are related by Poincare duality:
\be
\tilde H = e^{-\Phi}\, {\ast H}
\la{duality}
\ee
where $\ast$ denoted the Hodge dual. (Since this equation is conformally
invariant, it is not necessary to specify which metric is chosen in forming the
dual.) This ensures that the roles of field equations and Bianchi identities in
the one version of supergravity are interchanged in the other. As field
theories, each supergravity seems equally as good. In particular,
provided we couple them to an appropriate super Yang-Mills multiplet, then both
are anomaly-free\footnote{There are many anomaly-free groups in $D=6$ \cite{Salam}.}.

Now let us consider the above actions dimensionally reduced to $D=4$ and let the spacetime
indices be ${\mu,\nu}=0,1,2,3$.  The fundamental supergravity becomes
\[ 
I_4=\frac{1}{16\pi G}\int d^4x
\sqrt{-g}e^{-\eta}[R_g+g^{\mu\nu}\partial_{\mu}\eta\partial_{\nu}\eta-
\frac{1}{12}g^{\mu\sigma}g^{\nu\lambda}g^{\rho\tau}H_{\mu\nu\rho}H_{\sigma\lambda\tau}
\]
\be
-\frac{1}{2T_2{}^2}g^{\mu\nu}\partial_{\mu}T\partial_{\nu}\bar T
+...]
\la{sugra1} 
\ee
where the four dimensional fundamental metric is given by
$g_{\mu\nu}=G_{\mu\nu}=e^{\eta}g^c{}_{\mu\nu}$, where $g^c{}_{\mu\nu}$ is the four-dimensional
canonical Einstein metric.  The four-dimensional shifted dilaton $\eta$ is given by  %
\be
e^{-\eta}=e^{-\Phi}\sqrt{det G_{mn}}
\ee
and $T$ is the
modulus field of (\ref{T}).  This action is manifestly invariant under the $T$-duality of
(\ref{sl2zt}), with $\eta$, $g_{\mu\nu}$ and $B_{\mu\nu}$ inert.
Its equations of motion and Bianchi identities (but not the action
itself) are also invariant under the $S$-duality of (\ref{sl2zs}), with $T$ and $g^c{}_{\mu\nu}$
inert.  The axion field ${\rm a}$ is defined by
\be   
\epsilon^{\mu\nu\rho\sigma}\partial_{\sigma}{\rm a}=
\sqrt{-g}e^{-\eta}g^{\mu\sigma}g^{\nu\lambda}g^{\rho\tau}H_{\sigma\lambda\tau}
\ee
Similarly, the dual supergravity becomes
\[ 
\tilde I_4=\frac{1}{16\pi G}\int d^4x
\sqrt{-\tilde g}e^{-\sigma}[R_{\tilde g}+\tilde
g^{\mu\nu}\partial_{\mu}\sigma\partial_{\nu}\sigma-
\frac{1}{12}\tilde g^{\mu\sigma}\tilde
g^{\nu\lambda}\tilde g^{\rho\tau}\tilde H_{\mu\nu\rho} \tilde H_{\sigma\lambda\tau}
\]
\be
-\frac{1}{2S_2{}^2}\tilde g^{\mu\nu}\partial_{\mu}S\partial_{\nu}\bar S +...]
\la{sugra2}  
\ee
where the four dimensional dual metric is given by $\tilde g_{\mu\nu}=\tilde
G_{\mu\nu}=e^{\sigma}g^c{}_{\mu\nu}$.  The modulus field $\sigma$ is given by
\be
e^{-\sigma}=e^{\Phi}\sqrt{det \tilde G_{mn}}
\ee
and $S$ is
the axion/dilaton field of (\ref{S}).  This action is manifestly invariant under the $S$-duality of
(\ref{sl2zs}), with $\sigma$, $\tilde g_{\mu\nu}$ and $\tilde B_{\mu\nu}$ inert.  Its equations of
motion and Bianchi identities (but not the action itself) are also invariant under the $T$-duality
of (\ref{sl2zt}),  with $S$ and $g^c{}_{\mu\nu}$
inert. The pseudoscalar modulus field ${\rm b}$ is defined by 
\be
\epsilon^{\mu\nu\rho\sigma}\partial_{\sigma}{\rm b}=
\sqrt{-\tilde
g}e^{-\sigma}\tilde g^{\mu\sigma}\tilde g^{\nu\lambda}\tilde g^{\rho\tau}
\tilde H_{\sigma\lambda\tau}
\ee
Thus we see already at the level of supergravity that the roles of $S$ and $T$ have traded places!

This trading of axion/dilaton and
moduli fields was also noted by Binetruy \cite{Binetruy} who, inspired by $D=10$ string/fivebrane
duality, compared the fundamental and dual supergravities obtained by compactification from $D=10$
on $T^6$. However, his choice of $T$ was different from ours: $e^{\sigma}=\sqrt{det G_{mn}}$,
$m,n=4,5,6,7,8,9$ and ${\rm b}= B_{45}=B_{67}=B_{89}$ leading to a factor of $3$ in front of the
last term in (\ref{sugra1}).  As Binetruy points out, however, this choice has several unfortunate
consequences: (1) The factor of $3$ destroys the symmetry between $S$ and $T$; (2) It leads to
problems in formulating the dual supergravity in superspace  (see also \cite{Dauria}); (3) There is
no regime for which the fivebrane $\sigma$-model is weakly coupled.  This same choice of $T$ variables
was used by Sen and Schwarz \cite{Schwarz1,Sen7} when they concluded that ``...if there exists a dual
version of string theory for which the perturbative spectrum is manifestly $SL(2,Z)$ invariant, it
must be a theory of fivebranes".  If one replaces their variables by ours one arrives at the same
conclusion but with {\it fivebranes} replaced by {\it strings}. 

Furthermore, in unpublished work along the lines of \cite{Lu1}, Sen, Schwarz and the present author
tried and failed to prove that, for a fivebrane compactified on $T^6$, $SL(2,Z)_S$ is a symmetry that
interchanges the roles of the fivebrane worldvolume Bianchi identities and the field equations for
the internal coordinates $y^m$ $(m=4,5,6,7,8,9)$.  A similar negative result was reported by Percacci
and Sezgin \cite{Percacci}. In section \ref{sandt}, however, we shall prove that, for a dual string
compactified on $T^2$, $SL(2,Z)_S$ is a symmetry that interchanges the roles of the dual string
worldsheet Bianchi identities and the field equations for the internal coordinates $y^m$ $(m=4,5)$.  

These provide yet more reasons for preferring
a $D=6$ string/string duality explanation for $SL(2,Z)_S$ over a $D=10$ string/fivebrane duality
explanation, at least in the version where all six compactified dimensions and all six dimensions
of the worldvolume are treated on the same footing \cite{Schwarz1,Sen7}. In a future publication we
shall consider a modified version in which four of the six worldvolume dimensions wrap around four of
the six compactified dimensions.

\section{The fundamental string and the dual solitonic string}

The bosonic action of the fundamental $D=6$ heterotic string is given by
\be
 S_2 = {1 \over 2\p \ap} \int_{M_2} d^2 \x \left(
-\frac{1}{2}\sqrt{-\gamma}\gamma^{ij}\partial_iX^{M}\partial_jX^{N}G_{MN}
-{1\over 2}
\e^{ij} \partial_iX^{M}\partial_jX^{N} B_{MN} +...\right)
\la{1b}
\ee
where $\x ^i$ ($i=1,2$) are the worldsheet coordinates,
$\gamma_{ij}$ is the worldsheet metric and $(2\p\ap)^{-1}$ is the string
tension $T$. The metric and 2-form appearing in $S_2$ are the same as those in the supergravity
action $I_6$ of (\ref{a}), whose form is in fact dictated by the vanishing of the $S_2$
$\beta$-functions. This means that, to this order in $\alpha'$, under the rescalings with constant
parameter $\lambda$: 
\bea
G_{MN}& \rightarrow &\lambda^2 G_{MN} \nn \\
B_{MN}& \rightarrow &\lambda^2 B_{MN} \nn \\
e^{2\Phi}& \rightarrow &\lambda^2 e^{2\Phi} 
\la{1c}
\eea
both actions scale in the same way
\bea
I_6& \rightarrow &\lambda^2 I_6 \nn \\
S_2& \rightarrow &\lambda^2 S_2
\la{1d}
\eea
The combined supergravity-source action $I_6+S_2$ admits the singular
{\it elementary} string solution \cite{Dabholkar}
\[
ds^2= (1-k^2/r^2)[-d\tau^2+d\sigma^2 + (1-k^2/r^2)^{-2}dr^2 +r^2d\Omega_{3}{}^2]
\]
\[
e^{\Phi}=1-k^2/r^2
\]
\be
e^{-\Phi}*H_3=2k^2\epsilon_3
\la{fund}
\ee
where 
\be
k^2=\kappa^2 T/\Omega_3
\ee
$\Omega_3$ is the volume of $S^3$ and $\epsilon_3$ is the volume form.  It describes an
infinitely long string whose worldsheet lies in the plane $X^0=\tau,X^1=\sigma$.  This is now known
to be an exact solution  requiring no $\alpha'$ corrections \cite{Horowitz1}. Its mass per unit
length is given by %
\be
 M= T<e^{\Phi/2}>
\ee
and is thus heavier for stronger string coupling, as one would expect for a fundamental string.
The source-free action $I_6$ also admits the non-singular {\it solitonic} string solution
\cite{Lu3,Minasian} 
\[
ds^2= -d\tau^2+d\sigma^2 + (1-\tilde k^2/r^2)^{-2}dr^2 + r^2d\Omega_{3}{}^2
\]
\[
e^{-\Phi}=1-\tilde k^2/r^2
\]
\be
H_3=2\tilde k^2\epsilon_3
\la{sol}
\ee
whose tension $\tilde T$ is given by
\be
\tilde k^2=\kappa^2 \tilde T/\Omega_3
\ee
Its mass per unit length is given by
\be
\tilde {M}= \tilde T <e^{-\Phi/2}>
\ee
and is thus heavier for weaker string coupling, as one would expect for a solitonic string.
Thus, as promised in the Introduction, we see that the solitonic string differs from the
fundamental string by the replacements $\Phi\rightarrow -\Phi$, $G_{MN} \rightarrow \tilde
G_{MN}$, $H \rightarrow \tilde H=e^{-\Phi}*H$, $\alpha'\rightarrow \tilde \alpha'$.  The Dirac
quantization rule $eg=2\pi n$ ($n$=integer) relating the Noether ``electric'' charge 
\be
e=\frac{1}{\sqrt{2}\kappa}\int_{S^3}e^{-\Phi}*H_3
\ee
to the topological ``magnetic'' charge 
\be
g=\frac{1}{\sqrt{2}\kappa}\int_{S^3}H_3
\ee
translates into the quantization condition on the two tensions given in (\ref{Dirac1}).
Both the string and dual string soliton solutions break half the supersymmetries, both saturate
a Bogomol'nyi bound between the mass and the charge. These solutions are the extreme mass equals
charge limit of more general two-parameter black string solutions \cite{Horowitz2,Lu3}.

We now make the make the major assumption of this paper: the dual string may be regarded as
a fundamental heterotic string in its own right with a worldsheet action that couples to the dual
formulation of six-dimensional supergravity: 
\be
\tilde S_2 = {1 \over 2\p \tilde\ap} \int_{\tilde M_2} d^2 \tilde\x \left(
-\frac{1}{2}\sqrt{-\tilde\gamma}
\tilde\gamma^{ij}\partial_iX^{M}\partial_jX^{N}\tilde G_{MN}
-{1\over 2} \e^{ij} \partial_iX^{M}\partial_jX^{N}\tilde B_{MN} +...\right)
\la{1e}
\ee
where $\tilde \x^i$ ($i=1,2$) are the dual
worldsheet coordinates, $\tilde \gamma_{ij}$ is the dual worldsheet
metric and $(2\p\tilde \ap)^{-1}$ is the dual string tension $\tilde T$.
The metric and 2-form appearing in
$\tilde S_2$ are the same as those appearing in $\tilde I_6$ whose form will also be dictated by
the vanishing $\tilde S_2$ $\beta$-functions, assuming that this dual string admits a conformally
invariant formulation  Consistent with these identifications, we note that under the recalings with
constant parameter $\tilde \lambda$: %
\bea
\tilde G_{MN}&\rightarrow &\tilde \lambda^2\tilde G_{MN} \nn \\
\tilde B_{MN}&\rightarrow &\tilde \lambda^2 \tilde B_{MN} \nn \\
e^{-2\Phi}&\rightarrow &\tilde \lambda^{2}e^{-2\Phi} 
\la{1f}
\eea
both actions again scale in the same way:
\bea
\tilde I_6&\rightarrow &\tilde \lambda^2 \tilde I_6 \nn \\
\tilde S_2&\rightarrow &\tilde \lambda^2\tilde S_2
\la{1g}
\eea
The duality relation (\ref{duality}) is invariant under both the $\lambda$ and $\tilde \lambda$
rescalings.
  
It follows that the dual supergravity-source action $\tilde I_6 +\tilde S_2$ admits the dual string
(\ref{sol}) as the fundamental solution and the fundamental string (\ref{fund}) as the dual
solution.  When expressed in terms of the dual metric, however, the former is singular and the
latter non-singular.
  
\section{$S$-duality and $T$-duality}
\la{sandt}

In this section we derive $S$-duality as a symmetry of the dual string worldsheet by showing
that the equations of motion and Bianchi identities of $\tilde S_2$ are invariant under
$SL(2,Z)_S$.  On compactifiction to four dimensions, with $x^4,x^5=y^1,y^2$ the string Lagrangian
becomes 
\be
{\cal L}={\cal L}_x +{\cal L}_y + ...
\ee
where
\be
{\cal L}_x = -\frac{1}{2}\sqrt{-\gamma}\gamma^{ij}\partial_iX^{\mu}\partial_jX^{\nu}g_{\mu\nu}(X)
          -\frac{1}{2}\e^{ij} \partial_iX^{\mu}\partial_jX^{\nu} B_{\mu\nu}(X)
\ee
and
\be 
{\cal L}_y = -\frac{1}{2}\sqrt{-\gamma}\gamma^{ij}\partial_iy^{m}\partial_jy^{n}G_{mn}(X)
           -\frac{1}{2}\e^{ij} \partial_iy^m\partial_jy^n B_{mn}(X)
\la{L1} 
\ee
whereas the dual string Lagrangian becomes
\be
\tilde {\cal L}= \tilde {\cal L}_x +\tilde {\cal L}_y + ...
\ee
where
\be
\tilde {\cal L}_x=-\frac{1}{2}\sqrt{-\tilde\gamma}
\tilde\gamma^{ij}\partial_iX^{\mu}\partial_jX^{\nu}\tilde g_{\mu\nu}(X) 
               -\frac{1}{2} \e^{ij} \partial_iX^{\mu}\partial_jX^{\nu}\tilde B_{\mu\nu}(X)
\ee
and
\be 
\tilde {\cal L}_y =-\frac{1}{2}\sqrt{-\tilde\gamma}\tilde\gamma^{ij}
                    \partial_iy^{m}\partial_jy^{n}\tilde G_{mn}(X)
                   -\frac{1}{2} \e^{ij} \partial_iy^{m}\partial_jy^{n}\tilde B_{mn}(X)
\la{L2} 
\ee
Clearly, deriving $SL(2,Z)_S$ as a target space duality for the dual string is equivalent to
deriving $SL(2,Z)_T$ as a target space duality for the fundamental string.  This $T$-duality is a
well-known result, of course, but we repeat the proof here in order to focus on the subgroup given
by (\ref{sl2zt}). We follow the method given in \cite{Duff1} which involves the introduction of
dual $\sigma$-model action with compactified coordinates $\tilde {y}_m$ for which the roles of the
$y^m$ {\it worldsheet} field equations and Bianchi identities are interchanged: 
\be 
{\cal L}_{\tilde y} = -\frac{1}{2}\sqrt{-\gamma}\gamma^{ij}
     \partial_i\tilde {y}_{m}\partial_j \tilde {y}_{n}P^{mn}(X) 
                    -\frac{1}{2} \e^{ij} 
     \partial_i \tilde {y}_{m}\partial_j \tilde {y}_{n} Q^{mn}(X)
\ee
where $P^{mn}$ is the inverse of
\be
P_{mn}= G_{mn}+B_{mp}G^{pq}B_{qn}
\ee
and $Q^{mn}$ obeys
\be
P_{mn}Q^{pn}=B_{mn}G^{pn}
\ee
Let us define
\[
{\cal F}^{in} \equiv \sqrt{-\gamma} \gamma^{ij} \partial_j y^n
\]
\be
{\cal G}^{i}{}_n \equiv \sqrt{-\gamma} \gamma^{ij} \partial_j \tilde y_n
\ee
Then the field equations and Bianchi identities for $y^m$, or alternatively the Bianchi
identities and field equations for $\tilde y_m$, are given by 
\[ 
\partial_i \tilde {\cal G}^i{}_m=0
\]
\be
\partial_i \tilde {\cal F}^{im}=0
\la{y}
\ee
where 
\[
-\frac{\partial {\cal L}_y}{\partial y^m}=\tilde {\cal G}^i{}_m=G_{mn}  {\cal F}^{in} + B_{mn}
\tilde {\cal F}^{in} =\epsilon^{ij}\partial_j \tilde y_m
\]
\be
-\frac{\partial {\cal L}_{\tilde y}}{\partial {\tilde y}_m}=\tilde {\cal F}^{im}=P^{mn} {\cal
G}^{i}{}_n + Q^{mn} \tilde {\cal G}^i{}_n =\epsilon^{ij}\partial_j y^m
\la{fg}
\ee
The $T$-duality $O(2,2;Z)_T$ now follows by
showing that one transforms into the other under $O(2,2;Z)_T$ \cite{Duff1}.  In order to focus on
the $SL(2,Z)$ subgroup, however, it is more convenient to introduce
\[
*{\cal G}^{im} \equiv \epsilon^{mn} {\cal G}^i{}_n
\]
\be
* \tilde {\cal F}^i{}_m  \equiv \frac{1}{det G_{mn}} \epsilon_{mn} \tilde {\cal F}^{in}
\ee
which are also divergence-free:
\[
\partial_i *{\cal G}^{im}=0
\]
\be
\partial_i * \tilde {\cal F}^i{}_m=0
\ee
Now (\ref{fg}) may be written
\[
\tilde {\cal G}^i{}_m= G_{mn}{\cal F}^{im} -B_{mn}\epsilon^{np}* \tilde {\cal F}^i{}_p
\]
\be
* \tilde {\cal F}^i{}_m= -\frac{1}{det G_{mn}}\epsilon_{mn}P^{np}\epsilon_{pq}*{\cal G}^{iq}
                         +\frac{1}{det G_{mn}}\epsilon_{mn}Q^{np}\tilde {\cal G}^i{}_p
\ee
which, with the help of (\ref{T}), become
\be
\left(
\begin{array}{c}
* \tilde {\cal F}^i{}_m\\
\tilde {\cal G}^i{}_m
\end{array}
\right)
=\frac{1}{T_2}
\left(
\begin{array}{cc}
1 & T_1\\
T_1 & |T|^2
\end{array}
\right)
\left(
\begin{array}{c}
*{\cal G}^i{}_m/T_2\\
{\cal F}^i{}_m/T_2
\end{array}
\right)
\ee
where ${\cal F}^i{}_m \equiv G_{mn} {\cal F}^{in}$ and $*{\cal G}^i{}_m \equiv 
G_{mn}*{\cal G}^{in}$. In compact notation
\be
\left(
\begin{array}{c}
* \tilde {\cal F}\\
\tilde {\cal G}
\end{array}
\right)
= {\cal M}
\left(
\begin{array}{c}
*{\cal G}/T_2\\
{\cal F}/T_2
\end{array}
\right)
\ee
where
\be
{\cal M}=\frac{1}{T_2}
\left(
\begin{array}{cc}
1 & T_1\\
T_1 & |T|^2
\end{array}
\right)
\ee
Under (\ref{T}), ${\cal M}$ transforms as
\be
{\cal M}\rightarrow \omega^T{\cal M}\omega
\ee
where
\be
\omega=
\left(
\begin{array}{cc}
h& f\\
g & e
\end{array}
\right)
\ee
So, the combined $y^m$ field equations and Bianchi identities (\ref{y}) are indeed invariant under
$SL(2,Z)_T$ provided
\be
\left(
\begin{array}{c}
* \tilde {\cal F}\\
\tilde {\cal G}
\end{array}
\right)
\rightarrow
\omega^T
\left(
\begin{array}{c}
* \tilde {\cal F}\\
\tilde {\cal G}
\end{array}
\right)
\ee
and
\be
\left(
\begin{array}{c}
* {\cal G}/T_2\\
{\cal F}/T_2
\end{array}
\right)
\rightarrow
\omega^{-1}
\left(
\begin{array}{c}
* {\cal G}/T_2\\
{\cal F}/T_2
\end{array}
\right)
\ee
Finally, we have to show that the equations for $X^{\mu}$ and $\gamma_{ij}$ are also invariant.
The latter says that, up to an arbitary conformal factor, $\gamma_{ij}$ is just the induced metric
$h_{ij}$
\be
h_{ij}=\partial_iX^{\mu}\partial_jX^{\nu}g_{\mu\nu}(X)
        +\partial_iy^{m}\partial_jy^{n}G_{mn}(X)
\ee
Since $\partial_iX^{\mu}$ and $g_{\mu\nu}$ are inert under (\ref{T}), $h_{ij}$ transforms as :
\be
h_{ij} \rightarrow \partial_iX^{\mu}\partial_jX^{\nu}g_{\mu\nu}(X)
        +\partial_i{\tilde y}_{m}\partial_j{\tilde y}_{n}P^{mn}(X)= h_{ij} 
\ee
Since $\gamma_{ij}$ is invariant, it is not difficult to show that the $X^{\mu}$ equation is also.

We have thus proved the invariance of the fundamental string under $SL(2,Z)_T$. Comparing
(\ref{L1}) and (\ref{L2}), it follows, {\it mutatis mutandis}, that the dual string is invariant
under $SL(2,Z)_S$. Note that the supergravity action (\ref{sugra1}) is actually invariant 
under the continuous group $O(2,2)$; it is the toroidal nature of the compactification
that yields the discrete subgroup $O(2,2;Z)_T \sim SL(2,Z)_O \times SL(2,Z)_T$.  Similar remarks
apply to the dual action (\ref{sugra2}), and thus we learn that the discrete nature of the
$S$-duality group also has its origin in toroidal compactification of the dual string.

Since we are now dealing with four spacetime dimensions, it ought to be possible to describe both
the fundmental string and the dual string as elementary and solitonic solutions directly in
four dimensions. This is indeed the case.  The fundamental action (\ref{sugra1}) admits an as
elementary solution  \cite{Dabholkar,Sen5} the fundamental string
\[
ds^2=e^{\eta}(-d\tau^2+d\sigma^2)+dzd\bar{z}
\]
\be
S=a+ie^{-\eta}=\frac{1}{2\pi i}ln\frac{r}{r_0}
\la{string1}
\ee
where $z=x_3+ix_4$ corresponds to the transverse directions and $r=|z|$.  It also admits as a
soliton solution \cite{Khuri1} the dual string\footnote{This {\it dual} string of \cite{Khuri1} is
not to be confused with the {\it stringy cosmic string} of \cite{Greene}. The two solutions 
are different. The phrase {\it dual
string} has also been employed by Sen \cite{Sen5} to mean an $SL(2,Z)_S$ transform of the
fundamental string (\ref{string1}).  Both uses are valid, of course, but again should not be
confused.}  
\[
ds^2=-d\tau^2+d\sigma^2+e^{-\sigma}dzd\bar{z}
\]
\be
T=b+ie^{-\sigma}=\frac{1}{2\pi i}ln\frac{r}{r_0}
\la{string2}
\ee
It follows that the dual action (\ref{sugra2}) admits the the dual string as the elementary
solution and the fundamental string as the solitonic solution. Note that we may generate new
fundamental string solutions by making $O(2,2;Z)_{S}$ transformations on
(\ref{string1}) and new dual string solutions by making $O(2,2;Z)_{T}$ transformations on
(\ref{string2}).  So there is really an $O(2,2;Z)_{S}$ family of fundamental strings
\cite{Sen5} and an $O(2,2;Z)_{T}$ family of dual strings.  Once again, all this is consistent
with a duality of dualities.

\section{Subtleties with gauge fields}
\la{gauge}

In this section, we turn our attention to gauge fields which we have so far omitted from our
discussions.  These fall naturally into two categories: (1) the gauge fields already present in 
the $D=6$ string theory and whose details will depend on how we arrived at this theory; (2) the
$U(1)^4$ fields which arise in going from $6$ to $4$ dimensions on a generic $T^2$ and which
appear in $G_{\mu n}$ ( the Kaluza-Klein gauge fields) and $B_{\mu n}$ (the winding gauge fields).
 
We begin with (2) which are easier to discuss.  When these are included, the fundamental
supergravity action (\ref{sugra1}) becomes
\bea
\lefteqn{ I_4=\frac{1}{16\pi G}\int d^4x\sqrt{-g}e^{-\eta}[R_g +
g^{\mu\nu}\partial_{\mu}\eta\partial_{\nu}\eta
-\frac{1}{12}g^{\mu\lambda}
 g^{\nu\tau}g^{\rho\sigma}H_{\mu\nu\rho}H_{\lambda\tau\sigma}}\nn\\&&
-\frac{1}{4}g^{\mu\lambda}g^{\nu\tau}F_{\mu\nu}{}^a(LML)_{ab}
   F_{\lambda\tau}{}^b+ \frac{1}{8}
     G^{\mu\nu}Tr(\partial_{\mu}ML\partial_{\nu}ML)]
\la{sugra3}
\eea
where $F_{\mu\nu}{}^a=\partial_{\mu}A_{\nu}{}^a-\partial_{\nu}A_{\mu}{}^a$
and now $H_{\mu\nu\rho}=(\partial_{\mu}B_{\nu\rho}+2A_{\mu}{}^aL_{ab}F_{\nu\rho}{}^b)
+ {\rm permutations}$.  This action is invariant under $O(2,2)$ transformations with
\be
A_{\mu}{}^{a} \rightarrow \Omega^{a}{}_{b} A_{\mu}{}^{b}
\ee
Similarly, the equations of motion
continue to be invariant under $SL(2,R)$ transformations with 
\be
{\cal F}_{\mu\nu}{}^{a\alpha}
\rightarrow  \omega^{\alpha}{}_{\beta}{\cal F}_{\mu\nu}{}^{a\beta}
\ee
where $\a=1,2$ and
\be
{\cal F}_{\mu\nu}{}^{a1}=F_{\mu\nu}{}^{a}
\ee
and 
\be
{\cal
F}_{\mu\nu}{}^{a2}=S_2(ML)^a{}_{b}* F_{\mu\nu}{}^{b}+ S_1F_{\mu\nu}{}^{a}
\ee
Thus $T$-duality transforms Kaluza-Klein electric charge into winding electric charge (and
Kaluza-Klein magnetic charge into winding magnetic charge) but
$S$-duality transforms Kaluza-Klein electric charge into winding magnetic charge (and winding
electric charge into Kaluza-Klein magnetic charge).  In a way which should now be obvious, an
entirely similar story applies to the dual supergravity action (\ref{sugra2}) with $T$ and $S$
exchanging roles. (The same results could have been obtained from
the worldsheet point of view by including the background gauge fields in the calculations of the
previous section, but we shall omit the details.)  In this respect, string/string duality also
provides a stringy generalization of the old Montonen-Olive conjecture \cite{Montonen} of a duality
between the electrically charged particles of a fundamental supersymmetric theory and its
magnetically charged solitons. Indeed, the latter duality is in fact subsumed by the former in that
the solitonic magnetic {\it H-monopoles} \cite{Khuri5,Gauntlett} of the fundamental string are the
fundamental electric winding states of the dual string \cite{Khuri2,Rahmfeld}.  The Kaluza-Klein
states are common to both.

The four dimensional fundamental string solution (\ref{string1}) corresponds to the case where
all the above gauge fields have been set to zero.  As described in \cite{Sen2} a more general
solution with non-vanishing gauge fields may be generated by making $O(3,3)$ 
transformations on the neutral solution.  Such deformations are possible since the original
solution is independent of $x^0$ as well as $x^4$ and $x^5$.  However, since we want to keep
the asymptotic values of the field configurations fixed, this leaves us with an 
$O(2,1) \times O(2,1)$ subgroup. Not every element of this subgroup generates a new solution;
there is a an $O(2) \times O(2)$ subgroup that leaves the solution invariant. Thus the number of
independent deformations is given by the dimension of the coset space $O(2,1) \times O(2,1)/O(2)
\times O(2)$ which is equal to four, corresponding to the four electric charges of $U(1)^4$. An
exactly analogous statement now applies to the dual string solution (\ref{string2}).    

All of our discussions of the compactifying torus $T^2$ have so far assumed that we are at a
generic point in the moduli space of vacuum configurations and that the unbroken gauge symmetry
in going from $D=6$ to $D=4$ is the abelian $U(1)^4$.  However, we know that at special points
this symmetry may be enhanced \cite{Green}. For example if the radius of one of the circles lies at
the self-dual point
\be
R=\sqrt{\alpha'}
\ee
then the $U(1) \times U(1)$ becomes a $SU(2) \times SU(2)$. String/string duality now predicts a
new (non-perturbative) phenomenon, however.  An similar enhancement of the dual gauge symmetry can
also occur in the dual theory when
\be
R=\sqrt{\tilde \alpha'}
\ee
or, in other words, when Newton's constant obeys
\be
8G=\alpha'
\ee
Thus we learn that at special values of the dilaton and moduli vacuum expectation values, the string
and the dual string can have different\footnote{This is reminiscent of the
Goddard-Nuyts-Olive \cite{Goddard} phenomenon whereby the weight lattice of the fundamental gauge
group is given by the root lattice of the dual gauge group.} gauge groups! In general, the Narain 
\cite{Narain} mechanism applied to $T^2$ would yield $U(1)^2$ times any simply-laced gauge group of
rank $2$, namely $U(1)^2$, $SU(2) \times U(1)$, $SU(2) \times SU(2)$ or $SU(3)$.  Evidence from the
supergravity theories would then seem to indicate, however, that at these non-generic points the $S$
and $T$ dualities are no longer given by $O(2,2;Z)_T$ and $O(2,2;Z)_S$ since these are not preserved
by the non-abelian gauge interactions.

Finally we turn to the question we have been putting off so far, namely how the compactification
proceeded from $D=26$ (left movers) or $D=10$ (right movers) to $D=6$ and how to deal with the
corresponding gauge fields of type (1) above. This compactification might also be on a torus which
leads to a vector-like $N=2$ theory. Another choice, discussed in a recent paper on string/string
duality \cite{Minasian}, might go from $10$ to $6$ on a $K3$ manifold which leads to a chiral $N=1$
theory. In the latter case one can show in particular how to reproduce the Green-Schwarz spacetime
anomaly corrections to the $H$ field equations (a fundamental string one-loop effect) from the
Chern-Simons worldsheet anomaly corrections to the $\tilde H$ Bianchi identities (a dual string
tree-level effect), in accordance with the idea of interchanging the loop expansions.

Whatever compactification we choose, however, the simple picture that we have described so far no
longer obtains when we include the extra degrees of freedom involved in going from $D=26$ or $D=10$ to
$D=6$.  We lose the symmetry between the fundamental string and the dual string.  This is already
clear from the toroidal compactification: the fundamental string's target space duality group is now
enlarged to $O(6,22;Z)$ whereas that of the dual string remains $O(2,2;Z)$.  What kind of string would
the dual string be in general? Is it of the kind we already know? It is not even clear that it is
quantum mechanically consistent in the sense of having the right central charge required by conformal
invariance.  Perhaps a study of the zero modes of the dual string soliton in various
compactifications will throw light on these questions. Indeed, one may even entertain the idea
\cite{Minasian} that the requirement that the dual string be quantum mechanically consistent will
provide a non-perturbative mechanism for narrowing down the range of allowed superstring vacua. 

\section{Acknowledgements}

I would like to thank John Schwarz and Ashoke Sen for sharing their ideas on {\it duality of
dualities}, and Sergio Ferrara, Ramzi Khuri, Ruben Minasian, and Joachim Rahmfeld for useful
conversations.  I am grateful to the Director and Staff of the Isaac Newton Institute, and to the
organizers of the {\it Topological Defects} programme, for their hospitality.

\end{document}